# Purely equatorial lasing in spherical liquid crystal polymer microlasers with bipolar refractive index profile

*David Ripp, Nachiket Pathak, Vera M. Titze, Catriona A. Thomson, Andreas Mischok, Marcel Schubert**

D. Ripp, N. Pathak, V. M. Titze, Catriona A. Thomson, A. Mischok, M. Schubert
Humboldt Centre for Nano and Biophotonics, Department of Chemistry and Biochemistry, University of Cologne, Cologne, Germany
marcel.schubert@uni-koeln.de

V. M. Titze
Department of Sustainable and Bio-inspired Materials, Max Planck Institute of Colloids and Interfaces, Potsdam, Germany

Funding: This work received instrument funding by the Deutsche Forschungsgemeinschaft in cooperation with the Ministerium für Kunst und Wissenschaft of North Rhine-Westphalia (INST 216/1120-1 FUGG, 469988234), and research funding from the European Research Council under the European Union's Horizon Europe Framework Program (ERC Starting Grant agreement no. 101043047, HYPERION, to M.S.).

## Abstract

Liquid crystal whispering gallery mode (WGM) microlasers show high sensitivity to external stimuli and distinct spectral features, rendering them ideally suited for various sensing applications. They also offer intrinsic anisotropic optical properties, which can be used to shape and manipulate light even inside spatially highly symmetric structures. Here, we report the fabrication of a spherical bipolar liquid crystal polymer microlaser that tightly confines the optical path of WGMs to the equatorial plane. By controlled anchoring of the liquid crystal mesogens followed by polymerization, a fixed refractive index profile is formed within the spherical microcavity. Consequently, only transverse electric (TE) modes oscillating in the equatorial plane experience the high extraordinary refractive index, allowing to confine lasing into a single plane. Furthermore, we observe a characteristic splitting into sub-modes. By combining hyperspectral imaging, optical modeling, and theoretical calculations, we show that the observed sub-modes originate from the lifting of the energy degeneracy of azimuthal WGMs. The unique ability to confine lasing into a single plane makes these spherical microlasers independent of the exact position of the pump beam, which allows consistent localized sensing especially in combination with fast point scanning microscopes or inside highly dynamic biological environments.

## 1. Introduction

Whispering gallery mode (WGM) microlasers have emerged as a versatile tool for a broad range of sensing applications.[1,2] Isotropic solid microlasers with high sensitivity to changes in the external refractive index have been widely explored for biosensing.[1,3–7] Through surface functionalization, solid microlasers have enabled the detection of proteins and DNA molecules,[8,9] while their narrow linewidth emission has been exploited for cellular tagging and tracking.[5,10,11] Liquid crystal, elastomer, and oil droplet WGM microlasers provide an alternative sensing platform, detecting temperature,[12–16] electric fields,[17–19] mechanical forces,[20–22] biochemical molecules,[23–25] metal ions,[26] and surfactants.[27]

Due to their high symmetry, spherical WGM microresonators can support lasing along the inner circumference formed by the intersection of any plane that contains the center of the sphere. Therefore, the exact light path of the WGMs is often not fixed and will change depending on where the microresonator is pumped, and whether there are structures close to the surface that can interact with the WGMs. Consequently, WGMs can potentially reach any point on the surface of the resonator. This effect becomes apparent when employing scanned excitation schemes such as in light sheet microscopy[28] or confocal hyperspectral imaging,[29] where the WGMs change their oscillation plane depending on the position of the pump beam. It is also observed that light incoupling is most efficient at the rim of the microlasers while no or only very weak lasing is observed when the pump laser excites the center of the microspheres.[28]

A possible way to manipulate the light propagation inside of optical resonators is to modify their shape or to engineer the spatial distribution of the refractive index. The former was demonstrated in oil droplets and elastomer beads, where the deformation into ellipsoids induces symmetry breaking and mode splitting.[20,21] Manipulation based on the refractive index has most impressively been demonstrated in liquid crystal based microlasers[25,27] or in microbeads made from conjugated polymers with chiral morphologies.[30] In some cases, mode splitting into distinct modes has been reported in bipolar liquid crystal droplets,[27] yet without spatial localization of the individual modes within the microlaser or a detailed analysis about the origin of the splitting. Solid WGM microresonators with bipolar structures have also been demonstrated, but their photoluminescent emission spectrum lacks narrow WGM modes.[31,32] Moreover, spherical microlasers in which WGM confinement is restricted to a single plane have not been demonstrated to date. Thus, liquid crystal-based microlasers so far display the most versatile system to manipulate light propagation inside spherical microlasers. However, pure liquid crystal phases might be easily deformable or even undergo phase transitions, e.g. by changes of the surrounding temperature or surface-anchoring molecules.[18,25–27] Furthermore, for applications in biosensing it is also important to note that their liquid crystalline nature constitutes a significant risk given the recent assessment of the high cellular toxicity of many liquid crystal monomers.[33–35]

In this work, we introduce bipolar liquid crystal polymer WGM microlasers with low lasing thresholds. The microlasers show orientation-dependent lasing that depends on the relative orientation between the bipolar axis and the excitation beam. We further demonstrate that WGM lasing is only observed in the equatorial plane and that the corresponding lasing spectra have a distinct mode splitting. The spatial localization of these modes within the microlaser and the lifting of the mode degeneracy are explained by the combination of the anisotropic refractive index inside the microlasers and the distribution of the polar and azimuthal electric field components of the WGMs. Our results provide a detailed understanding of the interaction of light inside non-isotropic optical microresonators, while the unique lasing characteristics also provide new possibilities for localized biosensing.

## 2. Results and Discussion

To engineer the refractive index profile inside spherical microlasers, a liquid crystal polymer system was selected to combine anisotropic optical properties with the biocompatibility and mechanical advantages offered by a solid phase material. In this system, the molecular alignment of the mesogen-containing side chains of the polymer creates a direction-dependent refractive index, which will be the key to control light propagation in otherwise perfectly symmetric microspheres. We use rod-shaped acrylate-based monomers (**Figure 1b**) and bi-acrylate crosslinkers (**Figure 1c**) as the mesogens to form the liquid crystal phase, while polyvinyl alcohol (PVA) is used as a surfactant to induce planar anchoring, uniform alignment of the mesogens, and self-assembling into a bipolar configuration.[17,36] A subsequent crosslinking step (**Figure 1a**) arrests the molecular orientation and creates a polymer microbead with anisotropic optical properties. To achieve lasing, optical gain is introduced by adding strongly fluorescent dyes, either DCM (**Figure 1d**) or C545T **(Figure S1),** into the mesogen mixture. With this molecular system, we aim to manipulate the otherwise isotropic emission characteristics of spherical WGM microlasers to confine lasing to a single fixed plane (**Figure 1e**).

The first characterization of the optical anisotropy of the liquid crystal polymer is performed in thin films. Measurements of the refractive index (**Figure 1f**) of the non-aligned liquid crystal polymer with an ellipsometer yields the isotropic refractive index of 1.596 (at 600 nm), almost identical to the refractive index of isotropic polystyrene which is widely used as polymer matrix for microlasers.[5,10] Upon formation of a liquid crystal phase on a rubbed PVA film, the liquid crystal polymer exhibits pronounced optical anisotropy, with the ordinary and extraordinary refractive index measured to be $n_o$ = 1.511 and $n_e$ = 1.661 (both at 600 nm), respectively. The high extraordinary refractive index suggests that the liquid crystal polymer supports efficient confinement of light inside microspheres as small as 10 μm in diameter and below.[5,10]

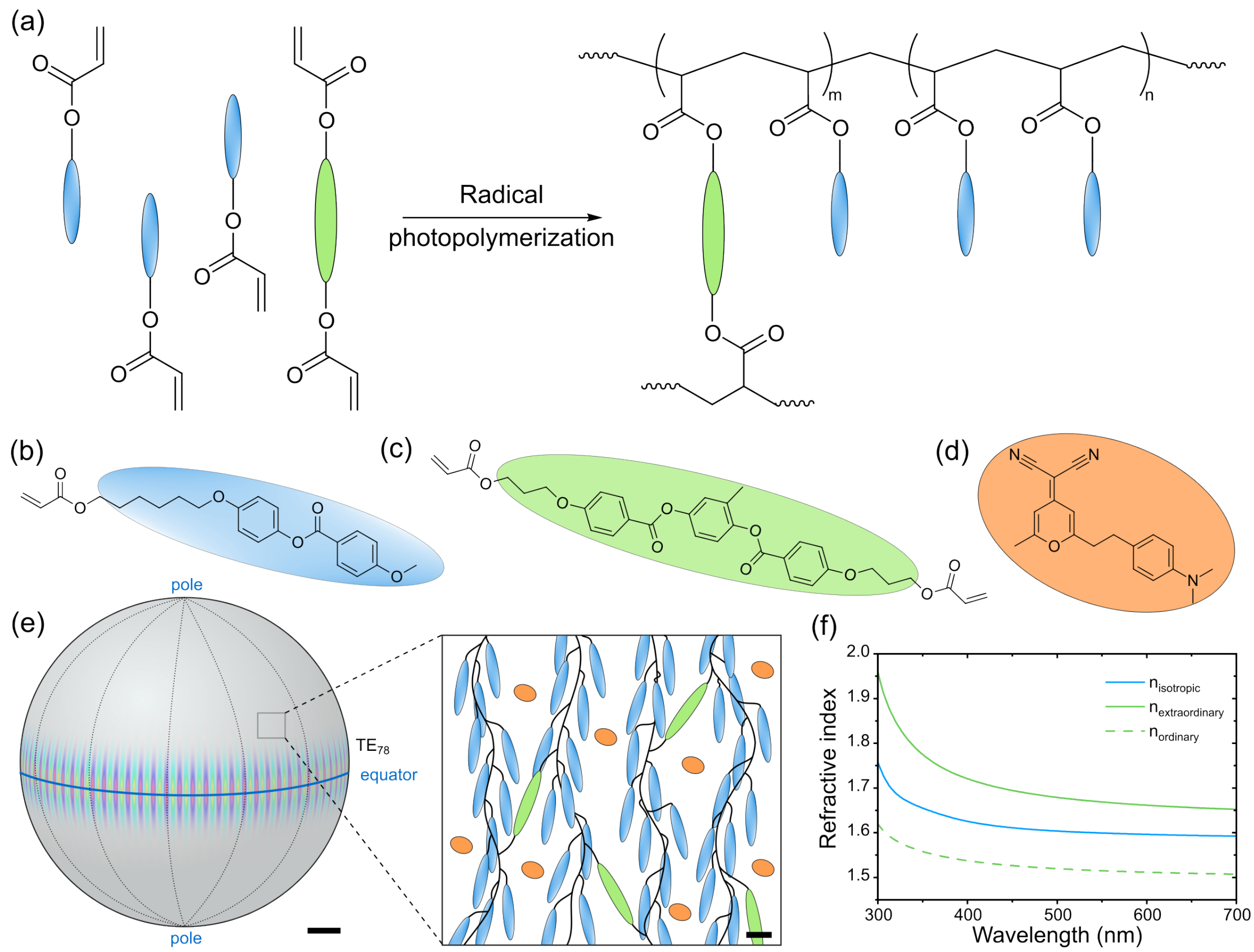

**Figure 1.** Schematic representation **(a)** of the radical photopolymerization between mesogens that create the polymer backbone and crosslinks between the chains. Molecular structures of **(b)** the monomer (blue), **(c)** the crosslinker (green) mesogens, and **(d)** the lasing dye DCM (orange). **(e)** Schematic illustration of a 10 µm large microlaser and a WGM (with transverse electric polarization and an angular mode number of $\ell$ = 78) confined to the equatorial plane. Scale bar, 1 µm. Inset: Illustration of the molecular structure of the polymer with aligned liquid crystal side chains. Dye molecules (orange) are embedded in the liquid crystal polymer matrix to provide optical gain. Scale bar, ~1 nm. **(f)** Measured ordinary (dashed green line), extraordinary (solid green line), and isotropic (blue line) refractive index curves of isotropic and aligned liquid crystal polymer films.

Monodisperse liquid crystal polymer microlasers are prepared using a focused flow droplet generator microfluidic chip with a nozzle size of 20 µm (**Figure 2a**). Molecular self-assembling into a bipolar liquid crystal phase within the droplets is achieved by adding PVA as a surfactant to the continuous phase.[17,36] The dispersed phase consists of a mixture of monomer, crosslinker, initiator, and lasing dye dissolved in dichloromethane. In the first step, droplets with diameters of approximately 25 to 30 µm are generated via pressure regulated flow of the two phases. Subsequent solvent evaporation causes the droplets to shrink and form pure liquid crystal droplets with diameters of approximately 9 to 10 µm (**Figure 2b**). Finally, ultraviolet (UV) light-induced crosslinking is used to fix the molecular alignment and to create solid liquid crystal polymer microlasers.

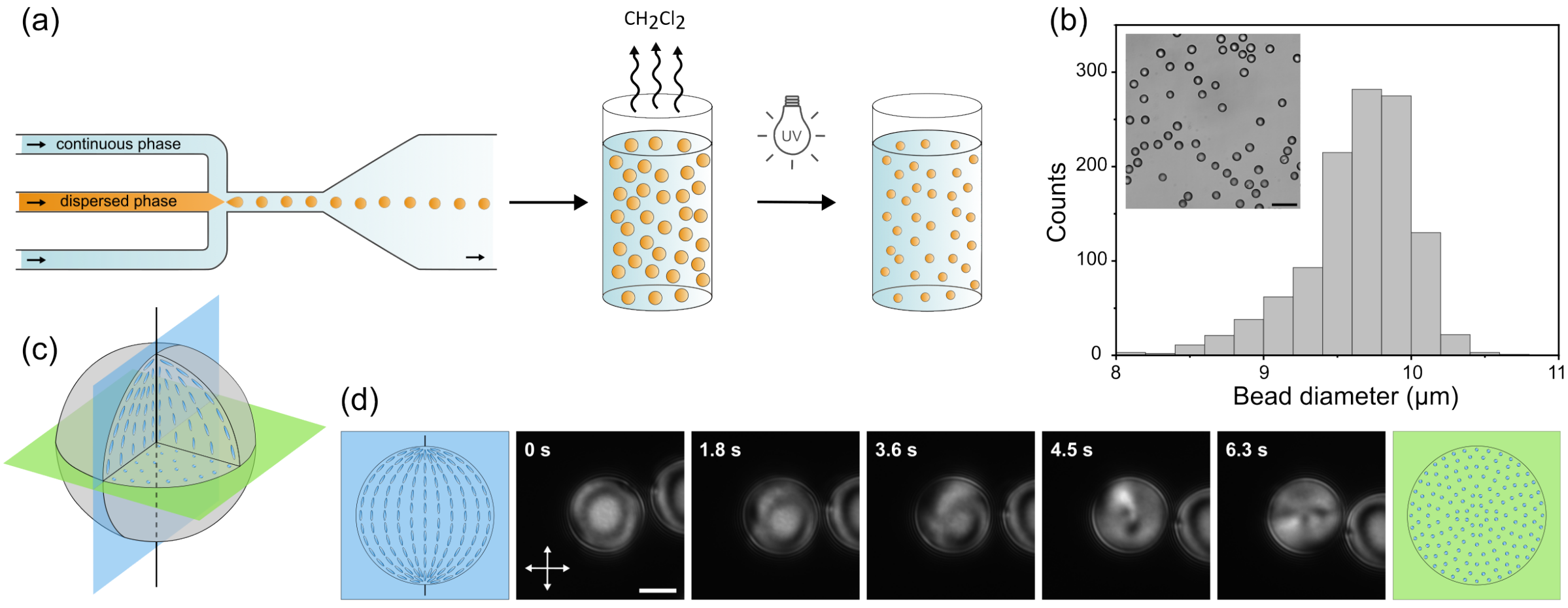

**Figure 2. (a)** Fabrication of liquid crystal polymer microlasers. Droplets of the dissolved mixture of the monomer, crosslinker, initiator, and lasing dye are prepared by microfluidics. The droplets are left until the solvent evaporates and subsequently polymerized by UV light to preserve the configuration of the liquid crystal molecules. **(b)** Typical size distribution of the resulting microlasers. Inset: Bright field microscopy image of the microlasers. Scale bar, 25 μm. **(c)** Schematic representation of the bipolar orientation of the mesogens inside the microlasers. Two cross-sections are shown, one through the plane containing the poles (blue) and one through the equatorial plane (green). **(d)** A series of cross-polarized images of a microlaser rolling inside water. The first and last microscopy image depicts the appearance of a bipolar microlaser in cross-polarized imaging when viewed sideways onto the bipolar axis or along the bipolar axis, respectively. For both cases, the orientation of the mesogens in the image is shown schematically.

The prepared microlasers exhibit a bipolar internal structure (**Figure 1e, 2c**) arising from the alignment of the mesogens. This bipolar configuration creates an anisotropic refractive index profile inside the microspheres, where the high extraordinary refractive index component is oriented along the great semicircles that run from one pole to the other, intersecting the equator at right angles (meridians). The orientation of this internal structure is confirmed experimentally by the observation of characteristic optical textures in cross-polarized images of a microbead freely rolling from one orientation to the other, transitioning from a side view of the bipolar axis to a top view along the bipolar axis (**Figure 2d**). Therefore, in cross-polarized images, a ring-like intensity distribution is observed for microlasers oriented with a side view onto the bipolar axis, whereas a cross-like or defect-like pattern is observed when the microlasers are viewed along the bipolar axis.[37,38]

The high refractive index of the liquid crystal polymer enables efficient optical confinement via total internal reflection, supporting WGMs even for microspheres with diameters as small as 8 µm **(Figure S2)** that are immersed in aqueous environments. Threshold measurements are performed to establish the typical pulse energy needed to generate WGM lasing. To avoid translational and rotational motion during the threshold measurements, microlasers are immobilized in a low concentration agarose gel. Emission spectra below and above threshold confirm the presence of WGMs (**Figure 3a**). A representative input-output curve for a 9.3 µm diameter microlaser in a sideways orientation is shown in **Figure 3b**, which has a threshold energy of 85 pJ. Statistical analysis of similarly oriented microlasers reveal an average threshold of 87 ± 21 pJ (**Figure 3c**). Also visible is a pronounced and almost equidistant splitting of the laser modes, which creates groups of 3-6 well-resolved modes, each about 300-400 pm apart (**Figure 3d**), and which progressively decrease in intensity (**Figure 3a,f**). We observe this well-resolved splitting for microlasers with a bipolar configuration and diameters ranging from 8 to roughly 11 µm, as well as for microlasers that contain a different fluorescent dye molecule **(Figure S1)**. Larger microlasers feature significantly more modes which makes the identification of regular mode spacings and patterns increasingly difficult **(Figure S2)**.

In addition, the emission intensity of the microlaser shows a strong polarization dependence with a contrast ratio of about 10:1 **(Figure 3e)**. The maximum emission intensity is observed when the linear polarizer in the emission path is parallel to the bipolar axis (i.e. perpendicular to the equatorial lasing plane), a correlation that is independent of the in-plane direction of the bipolar axis **(Figure S3b-d)**. Furthermore, all sub-modes have the same polarization-dependent emission characteristics **(Figure S3a,b)**. Additionally, upon changing the external refractive index, a collective red-shift of approximately 145 pm was observed for all lasing modes which implies that they comprise a comparable extend of the evanescent field as it is the case for the azimuthal mode family inside of spherical resonators (**Figure 3f**).[2,20,39] Based on these observations, we conclude that the microlaser emission is polarized parallel to the surface, consistent with the polarization of TE modes. These results indicate that for small bipolar microlasers immersed in an aqueous environment only TE but not TM modes are supported.

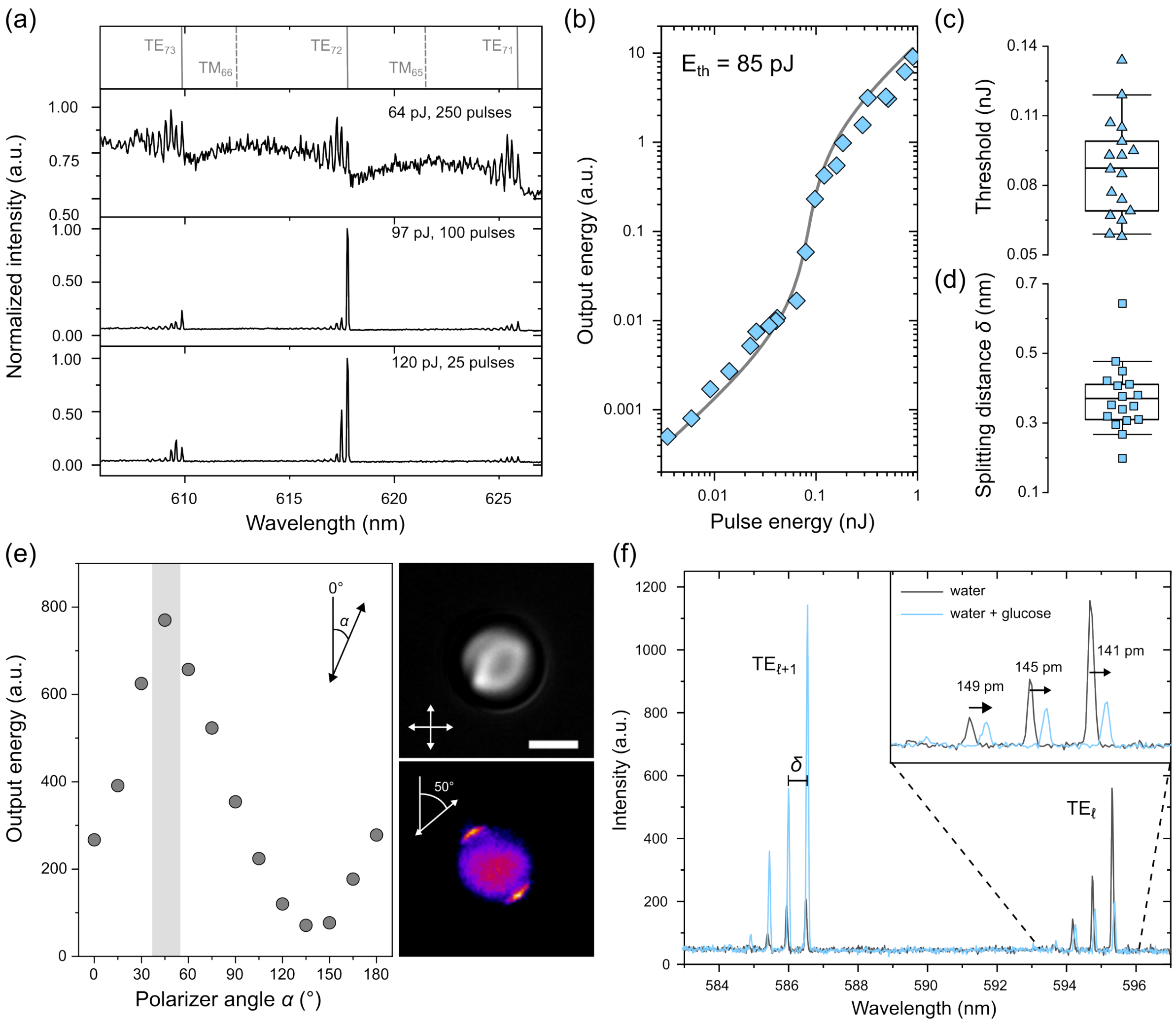

**Figure 3. (a)** Normalized lasing spectra at different pump energies below and above the laser threshold. The panel on top shows the calculated TE and TM mode positions for a spherical microresonator and the respective angular mode number. The bulk refractive index for the TE and TM modes was set to 1.66 and 1.51, respectively, corresponding to the expected values in the direction of the electric field oscillation. **(b)** Measured input-output characteristic (blue symbols) of a bipolar microlaser under nanosecond-pulsed excitation. Also shown is a fitted threshold curve (dark gray line), corresponding to a threshold energy of 85 pJ. **(c)** Distribution of measured laser thresholds of N=17 microlasers with diameters ranging from 9 to 10 µm. **(d)** Distribution of the splitting distance $\delta$ between two sub-modes of the same angular mode number, (see also (f)). The box plots show the mean and standard deviation, while whiskers represent the 10$^{th}$ and 90$^{th}$ percentile. **(e)** Polarization-dependent output energy of a microlaser showing a maximum at about 45°. The corresponding cross-polarized image (top) and the image of the microlaser emission above threshold (bottom) show the orientation of the microlaser with respect to the axis of the polarizer. The sketched angle of 50° indicates the direction of the bipolar axis extracted from the images. The WGMs are visible as two bright spots located on opposite sides of the circumference where the light is most efficiently collected by the objective. **(f)** Lasing spectra of a single microlaser measured in water (n=1.333) and in an aqueous

glucose solution (n=1,337), showing a collective shift of the sub-modes for both angular mode numbers ($TE_{\ell}$ and $TE_{\ell+1}$).

This conclusion is further supported by calculating the theoretical WGM spectrum using our previously described method.[3,29] Here, the expected free spectral range (FSR) ($\lambda_{TE,\,\ell} - \lambda_{TE,\,\ell+1}$) for a 9.3 μm large microsphere emitting around 610 nm and taking the measured anisotropic refractive index into account, is 8.0 nm, which is equal to the experimental FSR (**Figure 3a,f; S2**). In contrast, the expected difference between adjacent transverse electric (TE) and transverse magnetic (TM) modes ($\lambda_{TE,\,72} - \lambda_{TM,\,66}$) is about 4.8 nm, yet this mode spacing is not observed in the spectrum. Very similar characteristics are observed for other microlasers with diameters below 12 μm (**Figure S2**). In contrast, calculating WGM spectra assuming a homogeneous refractive index yields a FSR that is significantly larger than the experimentally observed value (**Figure S2**). The measured spectra are therefore consistent with the observed bipolar structure seen in cross-polarized imaging, supporting the assumption that the microlasers exhibit a strongly anisotropic internal refractive index profile.

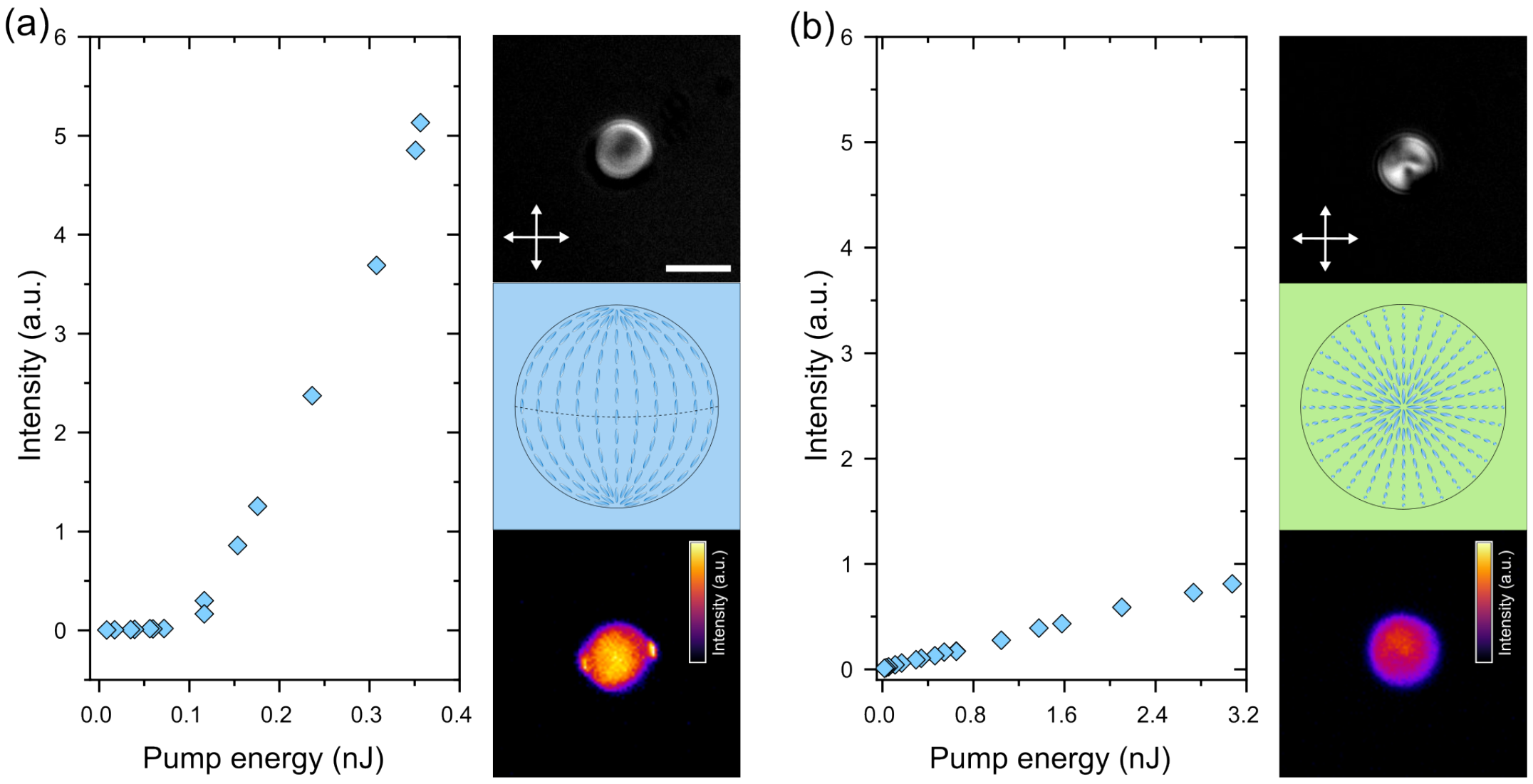

**Figure 4.** Threshold curves (left) and microscopy images (right) of microlasers oriented **(a)** sideways in relation to their bipolar axis, and **(b)** in a polar-facing orientation. For (a) and (b), the corresponding cross-polarized images (top), schematic representation of the liquid crystal orientation in a 3D sphere (middle), and a fluorescence microscopy image (bottom) are shown. The two bright spots in (a) indicate lasing in the equatorial plane while only homogeneous fluorescence is observed in (b). For pumping, a large homogeneous beam with a diameter of ~45 μm was used. Scale bar, 10 μm.

Next, we investigate how the lasing characteristics depend on the orientation between the pump beam and the bipolar axis. As before, the entire microlaser is excited by a homogeneous pump beam. First, we measure a microlaser lying sideways on the bipolar axis with the orientation information being inferred from the ring-like pattern in cross-polarized optical images (**Figure 4a**), which shows a lasing threshold of about 100 pJ. In contrast, for a microlaser oriented along the bipolar axis (**Figure 4b**), as inferred by the cross-like pattern, no lasing threshold is observed. Fluorescence images of the microlasers recorded at the maximum pump power also show a clear difference between the two orientations. In the sideways orientation, two bright spots appear on the opposite sides of the circumference of the microlaser. The anisotropy of the optical texture in the cross-polarized image allows us to identify the direction of the bipolar axis, revealing that these spots are located on the equator of the microsphere. The polar-facing orientation on the other hand only shows homogeneous fluorescence, even at pump energies up to thirty times higher than the threshold energy in the sideways orientation.

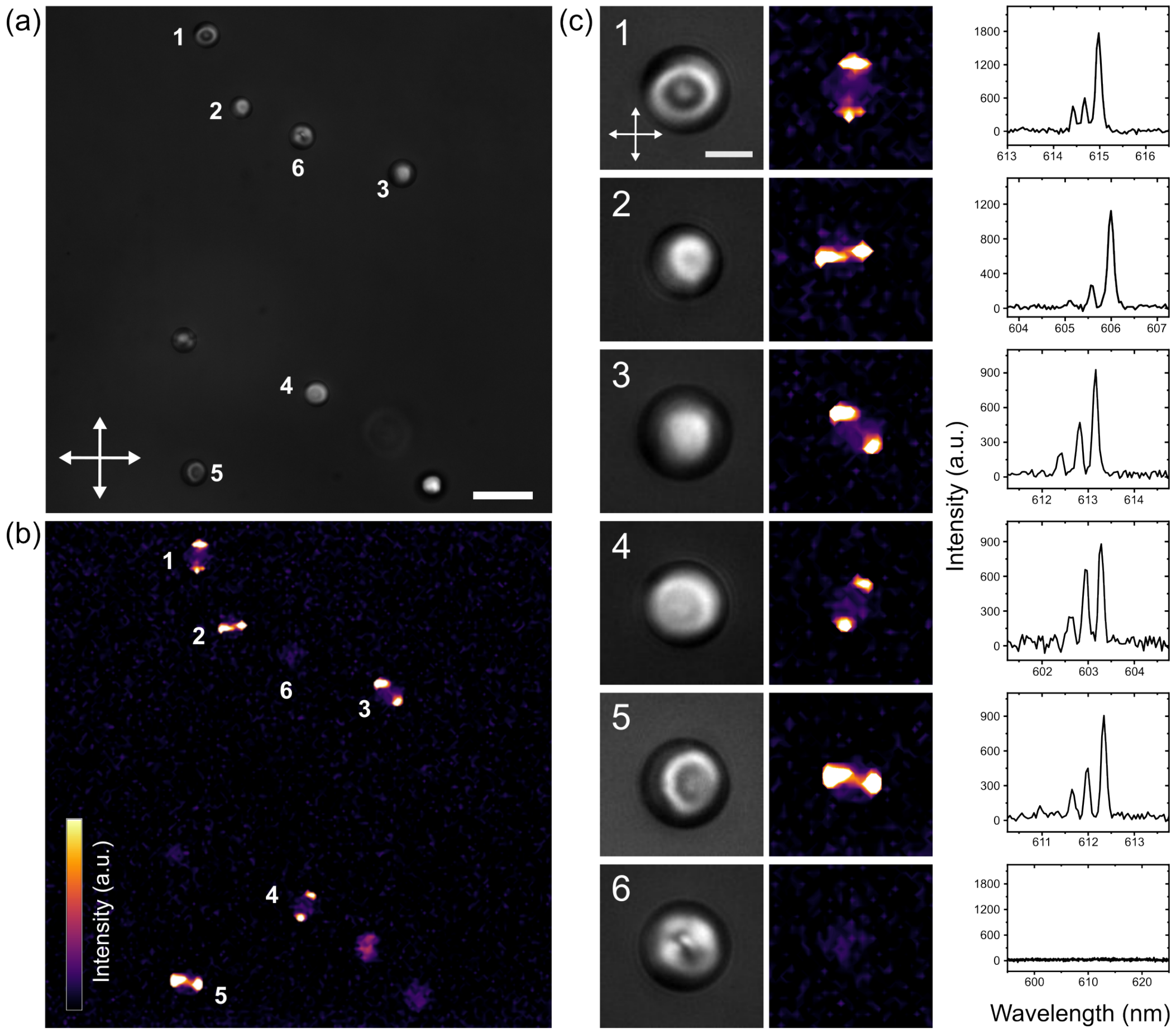

**Figure 5. (a)** Cross-polarized image showing multiple bipolar microlasers with different orientations. Scale bar, 25 µm. **(b)** Emission intensity map obtained by point scanning of the same field of view as in (a) with a confocal hyperspectral microscope at constant pump intensity (~0.4 nJ). **(c)** A zoomed-in cross-polarized image (left), intensity map (middle), and spectrum (right) of six selected microlasers

(labels corresponding to microlasers in a and b). The first five rows (1-5) show microlasers with a predominantly sideways orientation, each showing two characteristic bright spots on the equatorial plane, while the last microlaser is in a polar-facing orientation and shows almost no detectable emission. The spectra are taken from the pixel with the highest intensity. Scale bar, 5 μm.

To further investigate the effect of the relative direction between the pump beam and the bipolar axis, we measure a sample with randomly oriented microlasers (**Figure 5a**). We also change to a confocal point-scanning microscope which allows precise hyperspectral mapping of the emission of the microlasers.[29] Here, the field of view is scanned with 1 x 1 $\mu m^2$ spatial resolution and with the a focused pump laser set to a fixed energy (0.4 nJ), far above the previously established lasing threshold of microlasers in the sideways orientation. During scanning of the pump beam, a spectrum is taken at each point of the image. The resulting intensity map (**Figure 5b**) therefore reveals the exact points at which lasing can be excited. We observe that roughly half of the microlasers again show exactly two bright spots on their circumference which indicates that they are emitting in the lasing regime (**Figure 5b**), while the other half only shows very weak signals, a clear sign that they are still below the laser threshold. Analyzing cross-polarized images (**Figure 5c**) reveals that microlasers oriented roughly sideways on their bipolar axis (microlasers 1-5) display the characteristic spatial emission pattern as well as the previously observed splitting of the lasing modes into several sub-modes. In contrast, microlasers in a polar-facing orientation (microlaser 6) do not display any of the spectral or spatial characteristics associated with lasing. These experiments establish that only microlasers that have a predominantly sideways orientation can be excited above the laser threshold and only at two very distinct points on the circumference. This last observation is in strong contrast to our previous experiments where we scanned isotropic polystyrene microlasers which can be excited along their entire circumference **(Figure S4)**.[28,29] Thus, despite their spherical shape, WGM lasing of bipolar microlasers is confined purely to the equatorial plane.

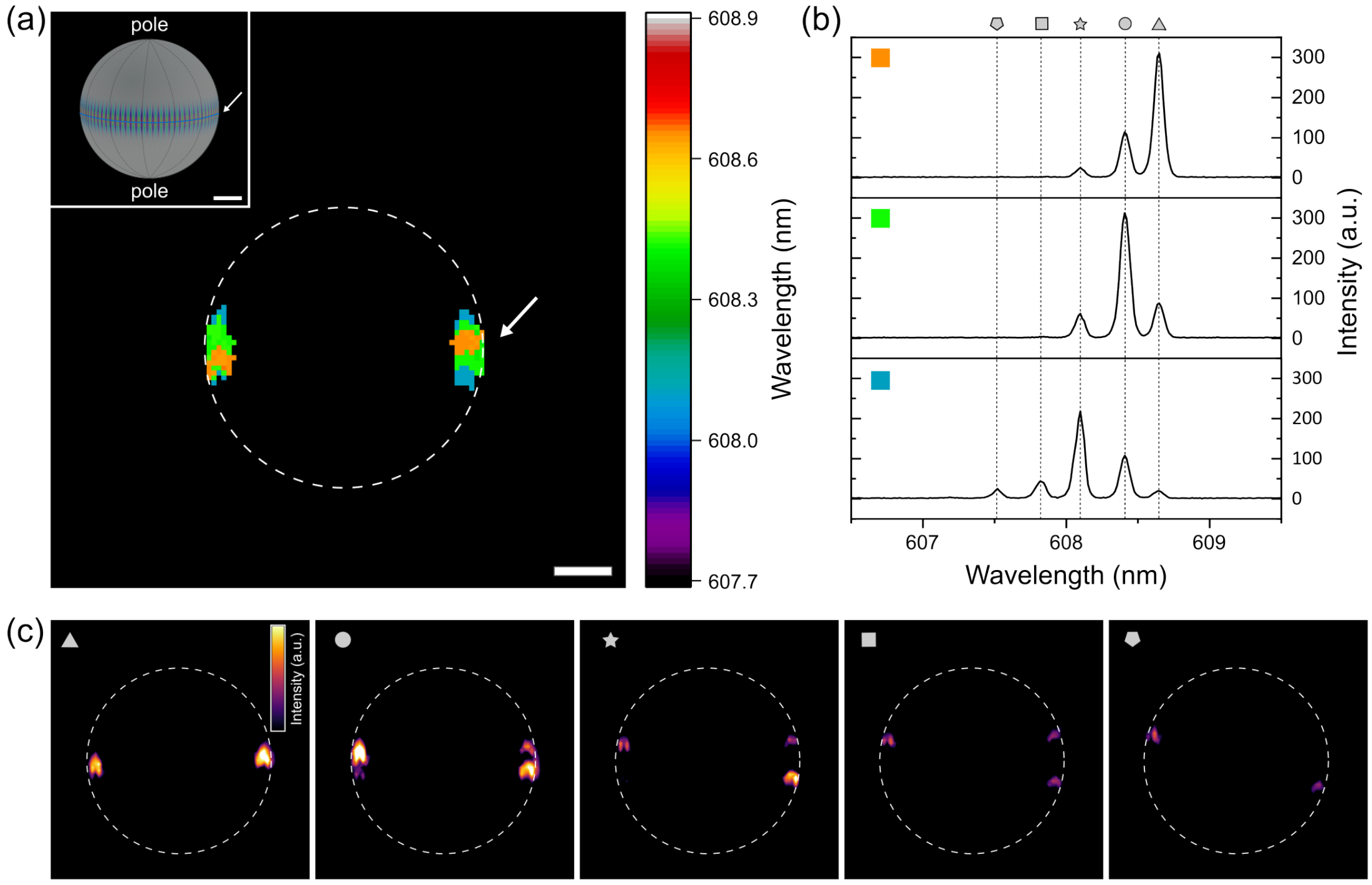

**Figure 6. (a)** High resolution map of the lasing signal of a single sideways-oriented bipolar microlaser obtained by point-scanning with a confocal hyperspectral microscope. For each pixel, the color represents the fitted central wavelength of the laser mode with the highest intensity. Two separate areas are visible, each showing a similar distribution of three distinct wavelengths. The position of the microlaser is represented by a dashed white circle. Inset: Schematic representation of the microlaser orientation. An arrow is pointing at the same position of the equatorial plane in the high-resolution map and the schematic representation. Scale bar, 2 µm. **(b)** Microlaser spectra calculated by averaging all pixels that have the same color in (a). Vertically dotted lines show that all spectra contain the same 5 modes, representing five sub-modes of a split TE mode which are located at 608.65 nm (triangle), 608.4 nm (circle), 608.1 nm (star), 607.8 nm (square), and 607.5 nm (pentagon). **(c)** Spectrally separated intensity maps of the five sub-modes visible in (b), showing the intensity distribution of a specific mode in the entire image (symbols corresponding to the mode assignment in (b)).

To identify the exact mode type, we perform high resolution confocal hyperspectral scanning of individual microlasers. As before, we observe lasing at the equatorial plane of a sideways-oriented bipolar microlaser (**Figure 6**, **Figure S5**). We then identify the dominant emission wavelength (i.e., the mode with the highest intensity) in each pixel and plot this wavelength in a color-coded map (**Figure 6a**). Two regions, each with three distinct lasing wavelengths, are observed at the characteristic edge position of the equatorial plane. Plotting the average emission spectra for each of the three colored areas reveals lasing sub-modes at 608.65 nm, 608.4 nm, 608.1 nm, 607.8 nm, and 607.5 nm (**Figure 6b)** in agreement with the characteristic splitting of the TE modes observed previously (**Figure 3a,f, 5c, S2, S3, S5**). Further analyzing the hyperspectral data, we plotted the intensity maps of the sub-modes (**Figure 6c**) which represent where in the image a specific sub-mode is emitting. Notably, all five lasing modes are spatially confined near the equatorial plane, but their position shifts progressively away from the equator, the lower the wavelength of the sub-mode is in the spectrum. The same characteristics are observed in the high-resolution scans of three other microlasers (**Figure S6**). Therefore, as the excitation position of the pump laser moves further away from the equatorial plane, modes with a shorter wavelength are excited preferentially. At the same time, the intensity of the sub-modes in the spectrum progressively decreases the further the mode extends towards the pole and the shorter the wavelength is, indicating a decreasing quality factor. We conclude that within a single group of modes, the sub-mode with the longest wavelength is purely confined to the equatorial plane, whereas modes with shorter wavelength are spatially offset towards the poles.

To classify the observed mode profiles (**Figure 6c**) we calculated the intensity distributions of azimuthal WGMs. The parameters were set to closely match the conditions of the microlaser shown in **Figure 6**, assuming modes with a TE polarization, an angular mode number of $\ell = 78$, and azimuthal mode numbers $m$ = 78, 77, 76, 75, and 74 (**Figure 7a**). The cross sections of these WGMs show that the intensity maxima of azimuthal modes are shifting from the equator towards the poles with decreasing mode number $m$, hereby forming an angle $\theta$ between the equatorial plane and the position of the maximum field intensity. This is consistent with the experimentally measured intensity maps (**Figure 6c**). Extracting $\theta$ from the intensity maps shows that the experimental values are in good agreement with the expected angles obtained from the calculated modes profiles **(Figure 7b)**, though an increasing deviation between the two is observed with increasing $\theta$. We attribute the difference to the limited resolution and low intensity in the hyperspectral maps and to possible distortions introduced by refraction of the light caused by the difference in refractive index of the microlaser and the surrounding water. Nevertheless, the characteristics of the experimentally observed mode patterns closely resemble those of azimuthal WGMs.

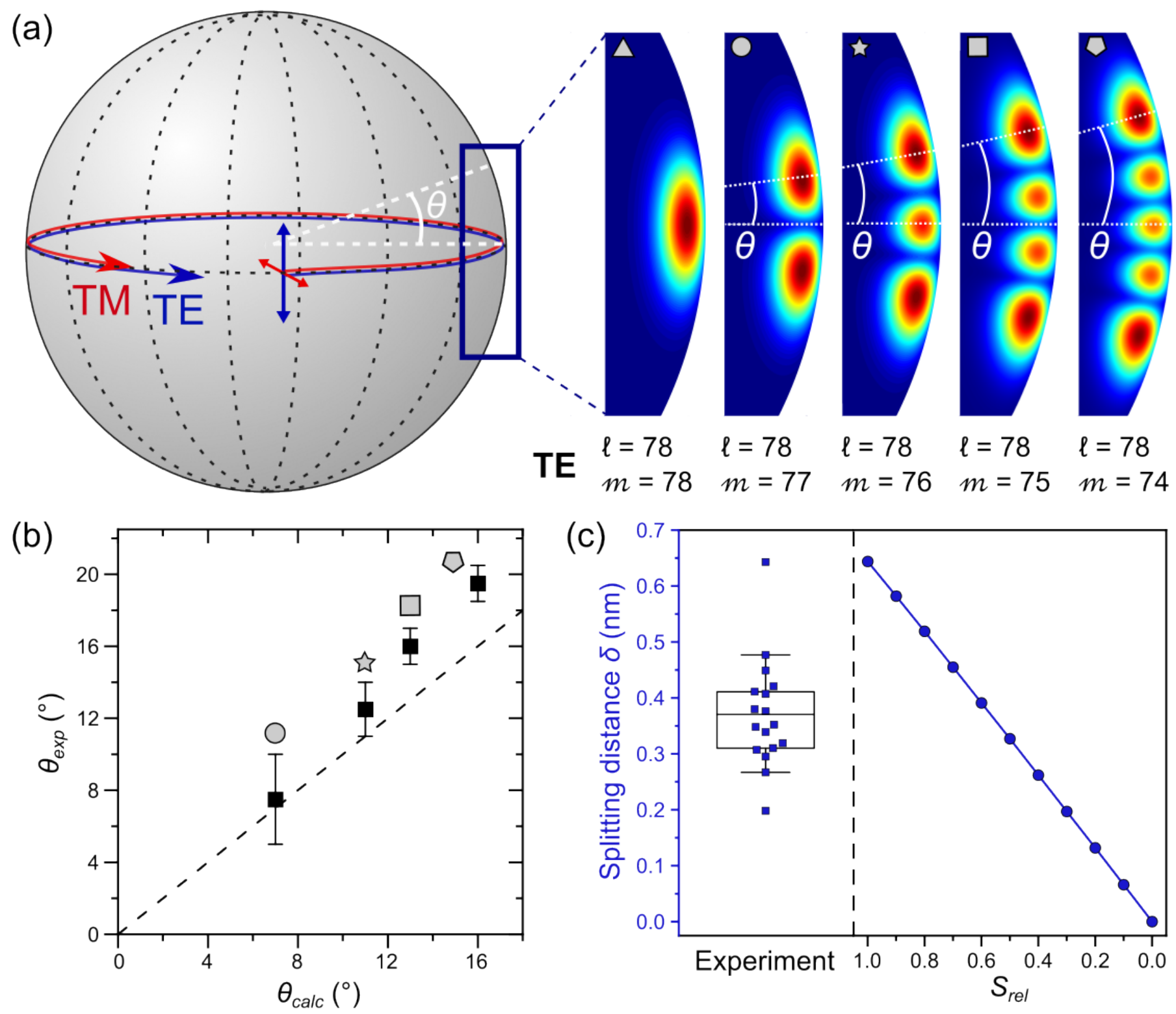

**Figure 7. (a)** Schematic representation of a bipolar microlaser with arrows showing the propagation and field orientation (i.e. polarization) of TE (blue) and TM (red) modes oscillating in the equatorial plane. The zoomed-in region around the equator shows the cross section of the intensity distribution of five WGMs with TE polarizations (TM modes very similar) with the same angular mode number ($\ell$ = 78) but different azimuthal mode numbers ($m$ = 78, 77, 76, 75, and 74). $\theta$ is the angle between the equator and the position of the maximum intensity of the mode in the polar direction. Note that $\theta$ increases with decreasing $m$, as the mode maximum shifts away from the equator and towards the poles. **(b)** Comparison of $\theta_{exp}$, obtained from the measured intensity maps in Figure 6, and $\theta_{calc}$, obtained from the calculated mode profiles shown in (a). The symbols correspond to the assignments of the modes in (a). The dashed line is the expected dependency. **(c)** Statistical distribution (left) of the splitting distance $\delta$ between the fundamental mode ($\ell$ = 78, $m$ = 78) and the next azimuthal mode ($\ell$ = 78, $m$ = 77) as shown already in Fig. 3d. Calculated splitting distance $\delta$ for different relative order parameters $S_{\mathrm{rel}}$ (right),

While azimuthal modes are energetically degenerate in perfectly spherical resonators, the observed splitting in energy implies that each mode experiences a different effective refractive index in bipolar microlasers. The origin of this effect is related to the boundary conditions of the electromagnetic field inside the sphere. Here, the electric field of TE-polarized modes oscillates parallel to the surface in the polar direction ($E_\theta$), but the sphere's curvature induces an additional field component in azimuthal direction ($E_\phi$) which increases for higher azimuthal modes (**Section 4.8**).[40–46] Because the electric field is not perfectly transversal to the direction of propagation they are also referred to as “quasi” TE modes. Due to the anisotropic refractive index inside the bipolar spheres, the polar and azimuthal field components will experience either the extra-ordinary or ordinary refractive index, respectively. We used the previously reported solutions for the electric field components of WGMs in spherical resonators to calculate the intensity of the WGMs in the polar and azimuthal directions.[40–46] Knowing the intensity distribution and the anisotropic refractive indices (**Figure 1f**) then allows to calculate the effective refractive index ($n_{\mathrm{eff}}$) of the azimuthal WGMs. This value is subsequently used to calculate the exact resonance wavelength of each azimuthal mode[39] which yields an equidistant mode splitting of $\delta$=650 pm between subsequent azimuthal modes (**Figure S7**). Independently of the theoretical analysis, numerical simulations are performed which yield a very similar splitting of $\delta$=745 pm (**Figure S8**). A possible origin for the slightly lower splitting obtained for the theoretical calculations might be the effect of the anisotropic refractive index which is generally not included in most theoretical descriptions, and which could lead to an underestimation of the azimuthal field component. Interestingly, the theoretical and simulated splittings are both larger than the average experimental splitting distance of $\delta$=370 ± 90 pm (**Figure 3d, 7c**). However, the degree of molecular alignment of the mesogens in the microscopic bipolar spheres is unlikely to be as good as in the aligned films where the values for the anisotropic refractive indices were obtained. Assuming that the order in the film is the best achievable for our liquid crystal elastomer system, we calculated the splitting for different relative order parameters $S_{\mathrm{rel}}$, which describes the average angular deviation of the liquid crystal molecules from the direction of the nematic director. This deviation will cause the molecules to twist their molecular axis away from the polar direction in which the TE modes oscillate predominantly. This will reduce the refractive index contrast between the polar and azimuthal direction and consequently reduce the splitting between different azimuthal modes. We find that a relatively high order parameter of about 0.6 produces a mode splitting comparable to the average experimental splitting distance (**Figure 7c**). We also perform experiments with liquid crystalline 5CB droplets, which were not polymerized into an elastomer matrix, hypothesizing that the higher refractive index anisotropy of 5CB and the improved order in the droplet compared to the elastomer should yield a larger splitting distance. The observed equidistant mode splitting is about 1.1 nm and very close to the theoretical value assuming an order parameter of 1 (**Figure S9**). In addition to the contributions from the refractive index anisotropy and the order parameter, we theoretically investigate how the deformation of the microresonator affects the mode splitting (**Figure S10**). We find that a relatively strong deformation of 200 nm would cause either an increase or decrease in the splitting distance by about 150 pm, depending on whether the spherical bead deforms into an oblate or prolate spheroid, respectively.

Finally, we also test the effect of the liquid crystal microlasers on the viability of cells. Fibroblast cells were incubated with the liquid crystal polymer microlasers as well as with 5CB droplets.[27] After two days, the liquid crystal polymer microlasers were still strongly fluorescent while the 5CB droplets dissolved within about 24h and induced cell death of almost the entire culture, indicating severe cytotoxicity (**Figure S11**). Although the experiment lacks statistical significance, it demonstrates the advantage of cross-linking the mesogens into the liquid crystal polymer matrix, which prevents the dissociation of the microlasers inside the cell culture medium.

## 3. Conclusions

In summary, we demonstrate the fabrication of monodisperse liquid crystal polymer microlasers with a spherical shape and a bipolar microstructure that confines lasing emission to a single, fixed plane. The microlasers have low lasing thresholds and are most efficiently excited when the pump beam is orthogonal to the bipolar axis. In contrast, excitation along the bipolar axis yields only fluorescence. The bipolar liquid crystal configuration induces an anisotropic internal refractive index profile that is experienced differently by modes with TE or TM polarization. In the small microlasers studied here, lasing is only observed for TE modes for which the polarization is (mostly) parallel to the high extraordinary refractive index of the mesogens. The resulting purely equatorial lasing was confirmed by confocal hyperspectral intensity maps correlated with cross-polarized optical images, which reveal both the microlaser orientation and the spatial localization of the lasing emission near the equator.

Furthermore, the bipolar microlasers show a pronounced and characteristic splitting of the WGMs which we resolved spatially and spectrally using high-resolution confocal hyperspectral scanning. Comparing measured and calculated mode cross sections allows to unambiguously classify the sub-modes as azimuthal WGMs. The bipolar refractive index profile that confines the modes also causes the lifting of the energy degeneracy of the azimuthal modes. Using the theoretical solutions for quasi-TE WGMs in spherical resonators we calculated the intensity of the electric fields in the polar and azimuthal directions and find that the ratio depends on the azimuthal mode number *m* (**Eq. 5**, **Sec. 4.8**). Subsequent calculations of the effective refractive index and exact resonance wavelength of the azimuthal modes predicts a homogeneous mode splitting which is consistent with the observed experimental splitting. Further experiments with 5CB droplet microlasers which we hypothesize to comprise an almost perfectly spherical shape and more ordered microstructure compared to a liquid crystal polymer, suggests that the splitting distance is directly connected to the molecular order inside the microspheres. Introducing an order parameter to describe the average angle of the mesogens relative to the bipolar director field, we show that the observed difference between the theoretical and measured splitting distances can be explained by a small degree of disorder of the mesogens inside the liquid crystalline polymer matrix (**Figure 7c**). Consequently, we attribute the variations in the splitting distances (**Figure 3d**) primarily to the variation in alignment efficiency between different microlasers. However, we note that elliptical deformations can also affect the mode splitting (**Figure S10**), although relatively large deformations of several hundred nanometers are required to significantly affect the mode splitting induced by the refractive index anisotropy. The splitting therefore provides an interesting way to analyze molecular order inside microscopic spheres, an information that is otherwise difficult to obtain.

With respect to the size-dependence of the laser emission, we show that microlasers up to about 11 µm in diameter show the desired purely equatorial lasing and regularly spaced sub-modes, while larger microlasers tend to have additional modes and less regular emission patterns (**Figure S2**). We attribute these more complex spectra to the onset of lasing of TM modes or higher order radial modes which is likely caused by the exponential increase in the radiative quality factor of the resonator with increasing diameter.[47] Furthermore, larger beads are more likely to develop defects and might be less organized, both effects reducing the anisotropy of the refractive index. However, the observed size-dependence does not limit the applicability of these microlasers as their size should be generally as small as possible for biological applications. Alternatively, optimizing the processing parameters and choosing different mesogens should allow to achieve the same lasing characteristics in larger microlasers.

Compared to isotropic spherical microlasers, the strictly confined equatorial lasing in spherical bipolar polymer microlasers displays a new emission characteristic of organic microlasers and offers unique advantages. For example, in a common hyperspectral scanning configuration, isotropic microlasers are scanned with a focused pump beam. Here, the moving excitation will constantly change the plane where WGMs oscillate because light incoupling is most efficient at the periphery of the microlasers (**Figure S4c**),[28,29] and because the plane of lasing must go through the center of the spherical resonator. If such scanning is performed in a heterogenous environment, e.g. a living cell or small animal, the laser spectra can show significant variations due to refractive index changes in the surrounding of the microlaser.[28,48] This variation in the resonance positions can then strongly limit the ability for barcoding which is one of the main applications of WGM microlasers. Similar problems arise for sensing applications where the mode shifts will rely on the exact pump position which is difficult to control within dynamic environments or deep inside tissue.[3]

Recently, light localization was reported in chiral polymer microlasers where directional emission of TE-polarized WGMs is observed in well-defined planes. However, due to the more complex microstructure, lasing in chiral microlasers occurs in almost all directions and doesn't involve azimuthal modes, rendering the emission characteristics different from the bipolar microlasers presented here.[30]

In contrast to isotropic or chiral microlasers, we envision that the light localization and unique spectral characteristics of bipolar microlasers offer exciting opportunities for spatially resolved biosensing, where the narrow sensing region can be aligned to a specific structural target inside the biological sample. The stable mesogen alignment within the spherical microlasers, together with the biocompatibility **(Figure S11),** and elastic properties of the material,[49] may also enable future applications in mechanical force sensing.[20] Here, the anisotropic structure of bipolar microlasers will most likely lead to an anisotropic elasticity of the polymer matrix.[50,51] Considering also the sensitivity of the sub-modes to changes in the geometry of the bipolar beads **(Figure S10)**, the single-plane emitting microlasers developed here might allow to decipher the direction from where the force acts on the microlaser. Moreover, the characteristic mode splitting might allow a more precise calculation of mechanical forces while the appearance of sub-modes in combination with the slight variation in the splitting distance between different microlasers could increase the number of uniquely identifiable barcodes. Finally, the here established connection between liquid crystalline order and spectral mode splitting also provides a promising way to optimize the fabrication protocols of the microlasers to synthesize microbeads with high optical and mechanical homogeneity. We therefore believe that microlasers with purely equatorial emission promise to unlock new applications in biosensing and barcoding of biological cells, or more generally, for sensing physical and mechanical parameters with microscopic optical sensors.

## 4. Methods

### 4.1. Materials and Chemicals

For microlaser fabrication, the monomer 4-methoxybenzoic acid 4-(6-acryloyloxy-hexyloxy)phenyl ester (CAS: 130953-14-9, ST03866, SYNTHON Chemicals) and the crosslinker 1,4-Bis[4-(3-acryloyloxypropyloxy) benzoyloxy]-2-methylbenzene (CAS: 174063-87-7, ST03021, SYNTHON Chemicals), the initiator 2-Benzyl-2-(dimethylamino)-4'morpholinobutyrophenone (Irgacure® 369, Tokyo Chemical Industry), and the liquid crystal 4′-Pentyl-4-biphenylcarbonitrile (5CB, CAS: 40817-08-1, Ossila) were used as received. For planar anchoring poly(vinyl alcohol) (PVA, n = approx. 2000, degree of saponification ca. 80mol%, Tokyo Chemical Industry) was used. As gain media 4-(Dicyanomethylene)-2-methyl-6-(4-dimethylaminostyryl)-4H-pyran (DCM, CAS: 51325-91-8, Sigma Aldrich) or 10-(benzo[d]thiazol-2-yl)-1,1,7,7-tetramethyl-2,3,6,7-tetrahydro-1H-pyrano[2,3-f]pyrido[3,2,1-ij]quinolin-11(5H)-one (C545T, CAS: 155306-71-1, LumTec) were used and solutions were made with dichloromethane ($CH_2Cl_2$) and toluene (Sigma Aldrich). All solutions were filtered by 0.22 µm PTFE filters (514-1275, VWR).

### 4.2. Microlaser preparation

Monodisperse microlasers were fabricated with commercial focused-flow droplet generator microfluidic chips (20 µm nozzle, 11002137, Micronit). For the dispersed phase, a mixture of monomer (50 mg), crosslinker (8.4 mg, 10 mol%), initiator (1 mg, 2 mol%) and lasing-dye (2 wt%, in relation to the total mass of the liquid crystal polymer mix) were dissolved in 1.2 ml of dichloromethane. For the continuous phase, a 0.2% w/v aqueous PVA solution was prepared. PVA is used here as surfactant to stabilize the emulsion and to induce planar anchoring and self-assembly of the mesogens in the droplets. Before using the prepared solutions for microfluidics, they were filtered by 0.22 µm PTFE filters. A 20 µm nozzle focused flow droplet generator chip was used, and pressure was applied by a custom build pressure system to control the flow of the two phases. The droplet size could be tuned between 25 and 30 µm. To evaporate the dichloromethane the collected droplets were shaken at 900 rpm for 2 h at room temperature. The resulting liquid crystal droplets had a monodisperse size between 9 and 10 µm. In a next step, the droplets were polymerized under gentle, continuous rotational movement of the whole emulsion, to prevent sedimentation, by exposing them to UV light (366 nm) for 20 min. The crosslinked liquid crystal polymer microspheres were separated from the surfactant solution by centrifugation and washed once with DI water. For storage, microspheres were re-dispersed in a 0.05% w/v aqueous PVA solution.

### 4.3. Refractive index measurement by ellipsometry

The refractive index was determined by variable angle spectroscopic ellipsometry (M2000, J.A. Woollam), modeled with a biaxial anisotropic oscillator model (CompleteEASE, J.A. Woollam). Measurements of aligned and isotropic thin films of the liquid crystal polymer were performed. For the aligned film, a silicon wafer was cleaned with acetone and isopropanol for 15 min in an ultrasonic bath each and spin coated with a 0.05% w/v PVA solution (3000 rpm, 1000 rpm/s, 60 s). The PVA film was rubbed manually several times in one direction with wipes (115-1803, VWR) and cleaned with toluene for 3 min in an ultrasonic bath. The same liquid crystal monomer, crosslinker, and initiator mix used for the microlaser preparation (but without the lasing dye) was dissolved in toluene with a concentration of 9% w/v. The dissolved mix was spin coated on the PVA film (900 rpm, 450 rpm/s, 30 s) and photopolymerized under UV light at room temperature for 20 min, achieving an aligned 600 nm thick liquid crystal polymer film. The film was characterized by ellipsometry to determine the ordinary and extraordinary refractive index. For the isotropic film, the liquid crystal polymer mix was directly spin coated on the silicon wafer without the PVA alignment film and polymerized for 20 min at 80°C under UV light.

### 4.4. Optical characterization

Lasing threshold measurements, refractive index sensing, and cross-polarized imaging were performed on a standard inverted fluorescence microscope (TE2000, Nikon). For the lasing experiments a green short-pulsed diode-pumped solid-state laser (515 nm, Flare NX 515-0.6-2, COHERENT), running at a repetition rate of 100 Hz with a pulse length of 1.15 ns, is coupled by a dichroic mirror into the microscope and directed on to the sample by a microscope objective (extra-long working distance (ELWD) 40x, numerical aperture (NA) 0.60, Nikon). Here, the entire microlaser is pumped with the green laser. The emission from the microlasers is collected by the same objective and directed to a spectrograph (Iso-plane SCT320, Princeton Instruments) equipped with a 1200 lines per mm grating and a CCD camera (Blaze 400HR, Princeton Instruments) and a spectral resolution of about 50 pm. Simultaneously, differential interference contrast (DIC) imaging or cross-polarized imaging can be conducted by using the transmission light pass of the microscope and by directing the transmitted light from the sample to a sCMOS camera (ORCA-Flash4.0 V3 C13440-20CU, Hamamatsu) using a dichroic mirror. For cross-polarized imaging, a polarizer is placed between sample and dia-illumination, and a second polarizer (analyser) is placed perpendicular to the polarization of the first polarizer between sample and camera. To induce flow and subsequent rolling of the microlasers during cross-polarized imaging, an abrupt thermal change was applied to the aqueous environment. Threshold measurements were performed with microlasers stabilized in a 0.25% w/v agarose matrix to avoid movement during the measurement. To change the pump energy, a set of optical density filters with different optical densities was inserted into the excitation light path before being coupled to the microscope. The polarization dependent output energy of the microlasers was measured by placing a polarizer (LPVISE2X2, Thorlabs) at different angles in the emission path before the spectrograph. Refractive index sensing was carried out with microlasers in water and consecutively adding a 30% w/v glucose solution to increase the refractive index.

### 4.5. Confocal hyperspectral imaging

Two different setups were used for confocal hyperspectral imaging. The spot size of the pump laser was below 1 µm in all hyperspectral confocal measurements. The first setup was used to scan a large field of view with multiple microlasers and simultaneous cross-polarized imaging (**Figure 5**). The liquid crystal polymer microlasers were fixed in a 0.25% w/v agarose matrix and imaged on a hyperspectral laser scanning confocal microscope. The setup was built around a standard inverted fluorescence microscope (TE2000, Nikon). A 515 nm pulsed laser (Origami-05XP, NKT Photonics) with a pulse width of 370 fs and a pulse repetition rate of 100 kHz was used as the pump source. The pulse energy was coarsely adjusted within the laser head, and a half-wave plate in combination with a polarizing beam splitter was used for fine control. The collimated laser beam was expanded and focused on the sample through a microscope objective (ELWD 40x, NA 0.60, Nikon). The pulse energy at the image plane was measured to be approximately 400 pJ. Fast galvo mirrors (5 mm aperture, Saturn 5B, ScannerMAX) were used to scan the beam across a 210 x 210 μm scan-field with a spatial resolution of 1 pixel/μm$^2$. The pixel dwell time was 1000 μs, consisting of 100 μs exposure and 900 μs spectrometer readout. The signal from the microlasers was collected through the same objective and de-scanned. A dichroic beam splitter and a focusing lens was used to direct the de-scanned signal to the entrance slit of the spectrograph (see above). The spectral CCD camera readout was limited to the top 50 lines in vertical binning mode to increase a spectral read-out rate of 1 kHz. The slit of the spectrometer served as a confocal pinhole and improved the spectral and special resolution. A full frame in this operation mode was captured in 44.1 seconds. The galvos and the spectrometer, microscope stage, and the fine laser power control were synchronized using a custom-built LabVIEW script. The pump laser was triggered by the spectrometer CCD to emit only during the exposure. In the acquired hyperspectral data set, each of the 44100 pixels in the scan-field contains a high-resolution spectrum which is analyzed by a custom python script to perform a Voigt fit on all lasing peaks to extract the peak wavelength, intensity, and widths. DIC and cross-polarized images of the same field of view were captured sequentially as described above to allow comparison of the microlaser orientation and axis of lasing.

The second hyperspectral microscope was used for high resolution scanning of single microlaser (**Figure 6**) by using a commercial laser scanning microscope (Stellaris 8, Leica), which we described in detail in a recent publication.[29] Here, a spatial (pixel) resolution of 190 nm and spectral resolution of 50 pm was achieved.

### 4.6. Mode calculation and visualization

Calculated resonances for the fundamental ($m = \ell$) angular WGMs were obtained using a custom code and the model proposed by Schiller[52] also used in a previous publication,[3] while for azimuthal modes the model by Gorodetsky[39] and Demchenko[53] was used, also applied in a recent publication.[20] All mode visualizations (**Figure 7**) were performed in MATLAB using a previously published toolbox which was further modified and extended for this work.[54]

### 4.7. Optical Simulations

Optical simulations of isotropic and bipolar microresonators were performed in COMSOL. Eigenmodes for isotropic and bipolar microresonators were simulated, using a 2D axisymmetric model and a fully defined permittivity tensor for the bulk material.

### 4.8. Quasi-TE mode and mode splitting calculations

WGMs in microspheres are classified as transverse electric (TE) and transverse magnetic (TM) modes only as an approximation because spherical boundary conditions prevent the field components from decoupling completely. As a result, quasi-TE modes are WGMs whose electric field is predominantly oscillating in the polar direction ($E_\theta$-dominated) yet still comprise a small azimuthal component ($E_\phi$) which oscillates in the propagation direction.[40–46] The exact vector spherical harmonics (assuming $E_r = 0$) of these two components are:

$$E_\theta(r,\theta,\phi) = E_0\,\psi_\ell(knr)\,\frac{1}{\sin\theta}\frac{\partial Y_{\ell m}(\theta,\phi)}{\partial\phi} \quad (1)$$

$$E_\phi(r,\theta,\phi) = -E_0\,\psi_\ell(knr)\,\frac{\partial Y_{\ell m}(\theta,\phi)}{\partial\theta} \quad (2)$$

$E_0$ is the electric field amplitude constant. $\psi_\ell(knr)$ is the Riccati-Bessel function describing the radial field profile. $Y_{\ell m}(\theta,\phi) = N_{\ell m}P_\ell^m(\cos\theta)e^{im\phi}$ is the spherical harmonic, $P_\ell^m(\cos\theta)$ is the associated Legendre function, and $N_{\ell m}$ a normalization constant. $\ell$ is the angular mode number and *m* the azimuthal mode number. To find the total integrated power ratio $\frac{I_\theta}{I_\phi}$ over the entire WGM, the squared magnitudes were integrated and solved resulting in (see Supporting Information):

$$I_\theta = \iint |E_\theta|^2 \sin\theta\ d\theta\, d\phi = 2\pi N_{\ell m}^2 m^2\left[\frac{1}{m}\frac{(\ell+m)!}{(\ell-m)!}\right] \quad (3)$$

$$I_\phi = \iint |E_\phi|^2 \sin\theta\ d\theta\, d\phi = 2\pi N_{lm}^2\left[\ell(\ell+1)\frac{2}{2\ell+1}\frac{(\ell+m)!}{(\ell-m)!} - m^2\frac{1}{m}\frac{(\ell+m)!}{(\ell-m)!}\right] \quad (4)$$

$$\frac{I_\theta}{I_\phi} = \frac{m}{\frac{2\ell(\ell+1)}{2\ell+1} - m} \quad (5)$$

The effective refractive index can be calculated by determining the exact normalized fractions of $I_\theta$ and $I_\phi$.

$$f_\theta = \frac{I_\theta}{I_\theta + I_\phi} \quad (6)$$

$$f_\phi = 1 - f_\theta \quad (7)$$

$$n_{eff}^2 = f_\theta n_e^2 + f_\phi n_o^2 \quad (8)$$

Finally, the exact resonance positions of the azimuthal quasi-TE modes were calculated with the model by Gorodetsky and Demchenko.[39,53]

$$\lambda = 2\pi n_{laser} a \left[ \ell - \alpha_q \left(\frac{\ell}{2}\right)^{\frac{1}{3}} + \frac{2p(a-b)+a}{2b} - \frac{\chi n}{\sqrt{n^2-1}} + \frac{3\alpha_q^2}{20}\left(\frac{\ell}{2}\right)^{-\frac{1}{3}} - \frac{\alpha_q}{12}\left(\frac{2p(a^3-b^3)+a^3}{b^3} + \frac{2n\chi(2\chi^2-3n^2)}{(n^2-1)^{\frac{3}{2}}}\right)\left(\frac{\ell}{2}\right)^{-\frac{2}{3}} + \left(\frac{\alpha_q^3+10}{1400} + \frac{(2p+1)^2 a^2 (b^2(1+3\mu)-a^2)}{32b^4}\right)\left(\frac{\ell}{2}\right)^{-1}\right]^{-1} \quad (9)$$

$a$ and $b$ are the equatorial and polar semiaxis, $\alpha_q$ the first Airy zero, $p = \ell - m$, $\chi$ the polarization parameter with $\chi = 1$ for TE modes and $\chi = 1/n^2$ for TM mode, $n$ the ratio of internal and external refractive indices ($n = n_{laser}/n_{external}$, with $n_{laser} = n_{eff}$, $n_{external} = 1.333$), and $\mu$ spheroid perturbation parameter.

### 4.9. Calculation of the order parameter

A relative order parameters $S_{\text{rel}}$ after the definition of Maier-Saupe[55,56] is calculated by:

$$S_{rel} = \left\langle \frac{3\cos^2\beta - 1}{2} \right\rangle \quad (9)$$

$S_{\text{rel}}$ describes the average angular deviation of the long molecular axis of the liquid crystal and the direction of the nematic director, described by the angle $\beta$. $S_{rel}$ is equal to one for a perfect alignment ($n_{eff} = n_e$). In our case we consider a relative alignment taking the experimental refractive index measurement ($n_e$ and $n_o$) of the aligned LCP film as highest possible order for alignment. For an isotropic orientation, $S_{rel}$ is equal to zero and for a perpendicular alignment orientation $S_{rel}$ is equal to $-\frac{1}{2}$ ($n_{eff} = n_o$).

For different $S_{rel}$, $n_e(S_{rel})$ and $n_o(S_{rel})$ are calculated as follow:[57]

$$n_e^2(S_{rel}) = n_{iso}^2 + \frac{2}{3} S_{rel}(n_{e,max}^2 - n_{o,min}^2) \quad (10)$$

$$n_o^2(S_{rel}) = n_{iso}^2 - \frac{1}{3} S_{rel}(n_{e,max}^2 - n_{o,min}^2) \quad (11)$$

with

$$n_{iso}^2 = \frac{n_{e,max}^2 + 2n_{o,min}^2}{3} \quad (12)$$

With the calculated refractive indices for different order parameters, the effective refractive index and the resulting splitting between fundamental and the next azimuthal mode can be calculated as described in **Section 4.8**.

### 4.10. Cell culturing

NIH 3T3 fibroblast cells were cultured in Dulbecco's Modified Eagle Medium supplemented with 10 vol% fetal bovine serum, 1 vol% Glutamax (100x) and 1 vol% penicillin-streptomycin (100x) in an incubator at 37°C, 5% $CO_2$ and 95% relative humidity. For cell experiments involving microdroplets and microlasers, $3 \times 10^4$ cells were seeded in 35 mm petri dishes (ibidi) and incubated for 24 h. The cells were then stained with 1 µl/ml Hoechst 33342, followed by a medium exchange after 1 h. Separate petri dishes were prepared for a control sample, 5CB microdroplets (produced via microfluidics) and liquid crystal polymer microlasers (with DCM dye as the gain medium). The 5CB microdroplets and the liquid crystal polymer microlasers were washed 3 times with sterile DPBS and 2 times with cell culture medium under sterile conditions. Subsequently, $2.5 \times 10^4$ 5CB microdroplets or liquid crystal polymer microlasers were added to the respective dishes. The control and the cell cultures containing 5CB microdroplets and polymer microbeads were additionally stained with 1 µl/ml propidium iodide to stain the nucleus of dead cells. Fluorescence and DIC imaging were performed on an inverted fluorescence microscope (TiE2, Nikon).

### Data availability

Data underlying the results presented in this paper will be made freely available.

### Conflict of Interest

The authors declare no conflict of interest.

### Supporting Information

Supporting Information is available from the Wiley Online Library or from the author.